\documentclass[letterpaper, 10 pt, journal, twoside]{ieeetran}

\IEEEoverridecommandlockouts

\usepackage{cite}   
\usepackage{amsmath,amssymb,amsfonts}
\usepackage{algorithmic}
\usepackage{graphicx}
\usepackage{textcomp}
\usepackage{subcaption}
\usepackage{xcolor}
\usepackage{todonotes}
\usepackage{soul}

\newcommand{\uvec}[1]{\boldsymbol{\hat{\textbf{#1}}}}

\usepackage{float}

\usepackage[font=footnotesize]{caption}

\usepackage{titlesec}
\titlespacing{\section}{0pt}{*0.35}{*0.35}
\titlespacing{\subsection}{5pt}{*0.3}{*0.3}
\IEEEoverridecommandlockouts                   

\usepackage[letterpaper, top=18.7mm, bottom=18.7mm, left=18.7mm, right=18.7mm]{geometry}

\author{Cristino de Souza Jr$^{1,2*}$, Rhys Newbury$^{3*}$, Akansel Cosgun$^{3}$, Pedro Castillo$^1$, Boris Vidolov$^1$, Dana Kuli\'{c}$^{3}$

\thanks{$^{1}$Univ. de Tech. de Compiègne, CNRS, Heudiasyc, CS 60319, France
        {\tt\footnotesize cristino.de-souza-junior@hds.utc.fr}}%
\thanks{$^{2} $Technology Innovation Institute, Abu Dhabi, UAE}
\thanks{Department of Electrical and Computer System Engineering, Monash University, Australia}%
\thanks{$^*$ Authors contributed equally to this paper.} 
\thanks{Digital Object Identifier (DOI): 10.1109/LRA.2021.3068952}
\thanks{\textcopyright2021 IEEE. Personal use of this material is permitted.  Permission from IEEE must be obtained for all other uses, in any current or future media, including reprinting/republishing this material for advertising or promotional purposes, creating new collective works, for resale or redistribution to servers or lists, or reuse of any copyrighted component of this work in other works.
}

}

\title{Decentralized Multi-Agent Pursuit using Deep Reinforcement Learning}

\begin{document}

\maketitle

\thispagestyle{empty}

\begin{abstract}
Pursuit-evasion is the problem of capturing mobile targets with one or more pursuers. We use deep reinforcement learning for pursuing an omnidirectional target with multiple, homogeneous agents that are subject to unicycle kinematic constraints. We use shared experience to train a policy for a given number of pursuers, executed independently by each agent at run-time. The training uses curriculum learning, a sweeping-angle ordering to locally represent neighboring agents, and a reward structure that encourages a good formation and combines individual and group rewards. Simulated experiments with a reactive evader and up to eight pursuers show that our learning-based approach outperforms recent reinforcement learning techniques as well as non-holonomic adaptations of classical algorithms. The learned policy is successfully transferred to the real-world in a proof-of-concept demonstration with three motion-constrained pursuer drones.
\end{abstract}


\section{Introduction}
Pursuit-evasion is the problem of capturing targets with one or more pursuers, with applications in robotics such as catching of a rogue drone or a ground target. 
With multiple-pursuers, decentralized systems are beneficial to avoid single points of failure. Classical algorithms for decentralized multi-agent pursuit~\cite{Janosov2017,Li2015,angelani2012collective} often assume omnidirectional pursuers, derive the local interaction rules from simple geometry and do not learn or adapt to evader behavior. For multi-agent teams consisting of wheeled robots or fixed-wing airplanes, the  non-holonomic kinematic constraints on the motion also need to be considered. To our knowledge, decentralized multi-agent pursuit subject to non-holonomic constraints has not been studied extensively by classical approaches in the literature.
Deep Reinforcement Learning (DRL) has also been successfully applied to multi-agent pursuit-evasion~\cite{Lowe2017, gupta2017cooperative, Huttenrauch2019, Xu2020}, however, most approaches to date did not consider real-world limitations such as local measurements and non-holonomic motion constraints and did not offer a thorough analysis of the system on operational metrics.

We propose a DRL approach to multi-agent pursuit. We consider a decentralized scenario in which non-communicating agents independently decide on their own actions based on local information. While our approach applies to any such system, in this paper, we focus on the specific scenario of capturing a finite speed but faster evader with multiple, non-holonomic pursuers in a bounded arena without obstacles. We treat pursuers as homogeneous agents and use shared experience to train a single policy executed independently by each agent at run-time. We use Twin Delayed Deep Deterministic Policy Gradient (TD3)\cite{fujimoto2018addressing}, a state-of-the-art DRL algorithm that was successfully applied to other domains\cite{matas2018simtoreal, chen2019modelfree}, with a state representation that encapsulates relative positional information of neighboring agents as well as the target and use a group reward structure that encourages good formations. During training, curriculum learning is applied to start with an easier version of the problem and gradually learn the task with increasing difficulty. In simulation experiments, we compare our approach to three state-of-the-art approaches, two classical methods and a DRL method. They are evaluated in terms of the capture rate and average timesteps to capture. We conduct further analysis on the effect of the number of agents, arena size, as well as using variable linear speed, curriculum learning, and formation score as part of the reward function. The trained policy is demonstrated in a proof-of-concept physical system with three pursuer drones subject to non-holonomic motion constraints.

The organization of this paper is as follows. After reviewing the relevant literature in Sec.~\ref{sec:related_work}, we define the problem of interest in Sec.~\ref{sec:setup}. Our multi-agent DRL method is presented in Sec.~\ref{sec:deep_rl}. We detail the experimental procedure in Sec.~\ref{sec:experiments} and present simulation results in Sec.~\ref{sec:results}. Finally, we describe the real-world drone implementation in Sec.~\ref{sec:implementation} before concluding with a brief discussion in Sec.~\ref{sec:conclusion}.

\section{Related Work}
\label{sec:related_work}

\subsection{Multi-agent pursuit}

Solutions to the pursuit-evasion problem either assume the `worst-case' adversary with infinite speed and complete awareness of the pursuers, or average-case behaviors~\cite{chung2011search}. Although single-agent pursuit-evasion is studied extensively in the literature~\cite{chung2011search,shneydor1998missile}, its extension to multi-agent systems still remains an open problem \cite{kamimura_ohira_2019}, of interest in biology~\cite{kamimura_ohira_2019}, physics~\cite{angelani2012collective, Janosov2017}, and engineering~\cite{Huang2011, Li2015}. 

Non-learning pursuit methods can be organized into deterministic and heuristic solutions. Deterministic methods attempt to solve the problem with traditional mathematics tools, such as pursuit curves analysis~\cite{shneydor1998missile} and differential games~\cite{Huang2011, Li2015}. Pursuit curve analysis formulates the trajectory of the pursuer analytically using differential equations. The system can then be solved to find the conditions of capture. Although the problem can be stated simply, closed-form analytical solutions are hard to obtain even using simplistic assumptions such as constant velocities and linear trajectories. Differential games formulate pursuit as a game, where the players must optimize an objective function, often the episode duration. However, it is challenging to define an appropriate objective function  with increasing problem complexity, such as with a larger number of agents or motion constraints. 

Heuristic solutions~\cite{Janosov2017, angelani2012collective, muro2011wolf} inspired by behavior-based decentralized approaches~\cite{vicsek1995novel}, use computational simulations that aim to find emergent group behavior based on local observation~\cite{vicsek1995novel, Olfati-Saber2006b}.
Angelani~\cite{angelani2012collective} proposed modeling up to a hundred autonomous pursuers as particles based on~\cite{vicsek1995novel}. Muro~\cite{muro2011wolf} proposed small-scale hunter strategies with up to 5 agents, arguing that the behavior observed in wolf-pack hunts can be simulated with simple rules. Janosov~\cite{Janosov2017} re-examined the concepts of the Vicsek particle model~\cite{angelani2012collective} in a more realistic scenario considering delays, accelerations, prediction and a faster target. However, these works assume access to information about the evader's position and velocity, which is not directly available from onboard sensors. 




The real-world applicability of many of the above approaches is limited due to their assumptions in observation and actuation. \cite{angelani2012collective} and \cite{Janosov2017} both consider an omnidirectional particle model, which is not directly transferable to many robotic platforms. In contrast, our observation model is expressed relative to each agent, which can be found directly using onboard sensors, such as LiDARs or depth cameras\cite{drone_detection}. We also consider a non-holonomic kinematic model, suitable for car-like mobile robots or fixed-wing airplanes.

\subsection{RL In Pursuit}
The pursuit-evasion game is a highly studied task in multi-agent RL~\cite{Desouky2011, Jouffe1998, Awheda2015}. However, most approaches apply only to omnidirectional agents, which cannot be easily transferred to real robotic applications without a loss in performance. Lowe~\cite{Lowe2017} presented an approach for multi-agent RL using an adapted version of an actor-critic algorithm extended to multi-agents. Their approach on the pursuit-evasion game with omnidirectional agents outperformed Deep Deterministic Policy Gradient (DDPG)~\cite{lillicrap2015continuous}. Xu~\cite{Xu2020} considered pursuit-evader games with non-holonomic agents, where new agents can join the game. They adapted Bi-directional Recurrent Neural Networks~\cite{schuster1997bidirectional} and DDPG. However, they only consider a situation with $3$ and $5$ agents. Furthermore, the observation also assumes global information about other agents, limiting the applicability to real-world situations. 
A few pursuit-evasion works consider non-holonomic constraints \cite{Huttenrauch2019,KOTHARI20141977}. Hüttenrauch~\cite{Huttenrauch2019} studied multi-agent pursuit-evasion systems by considering the agents as interchangeable and the exact number irrelevant. They create a new state representation based on mean embedding of distributions. Their work focuses on scalability and shows that their system can operate with up to fifty agents. 

\cite{UAVPursuit2, UAVPursuit,DronesChasingDrones} learn a policy directly from images in a pursuit-evasion scenario with one chaser and one evader. These policies are then successfully transferred to a real-world scenario and show good performance.

\subsection{Curriculum Learning}
Curriculum learning~\cite{bengio2009curriculum} is a learning paradigm to help improve speed of convergence and reduce local minima by gradually increasing the complexity of training data. This learning paradigm has been been widely used for RL\cite{portelas2020automatic, tidd2020guided} and deep learning~\cite{graves2017automated}, and shown to solve problems which were previously considered intractable~\cite{gupta2017cooperative}. 

Our work, while borrowing ideas from both classical and learning-based methods, focuses on using DRL to improve the pursuit performance and consider operational metrics such as capture success rate and the average time to capture. Furthermore, we propose a method that is suitable for sim-to-real policy transfer with realistic observation models and non-holonomic constraints. To our knowledge, we are the first to demonstrate a real-world pursuit-evasion implementation with multiple pursuers using a DRL policy.

\section{The pursuit-evasion scenario}
\label{sec:setup}
Our pursuit-evasion problem consists of multiple homogeneous, slower pursuers chasing a single, faster target. The goal for pursuers is to move as a group so that the freedom of the target is constrained to the point where one of the pursuers `captures' the target in the shortest time possible. We consider a trial successful if, at any point during the trial, the distance between the evader and at least one of the pursuers is less than a given collision radius ($d_{i,T} < d_{cap}$). If the target is not captured within a fixed time period $T_{timeout}$, then a timeout occurs, and the trial is considered unsuccessful. Collisions between pursuers do not result in a failure, however, it is discouraged within the reward function (Sec.~\ref{subsec:reward}), an important feature for collision avoidance in real-world implementation (Sec.~\ref{sec:implementation}). The pursuer motions are subject to non-holonomic kinematic constraints, while the evader is omnidirectional and thus not subject to such constraints. We adopt a unicycle model for each pursuer $i$:
\begin{subequations}
\begin{align}
    \dot x_i &= v \cos \psi_i \\
    \dot y_i &= v \sin \psi_i \\
    \dot \psi_i &= \omega
\end{align}
\label{evolution_model}
\end{subequations}
where ($x$,$y$) is the position, $\psi$ is the heading angle, $v$ is the linear velocity and $\omega$ is the angular velocity. We will first assume that all agents have a constant $v$ and the only controllable variable is $\omega$ with limits $\omega_{min} \leq \omega \leq \omega_{max}$, following the classical formulation approach\cite{Janosov2017}. Later we will relax this assumption to allow the pursuers to vary both their angular and linear velocities. We approximate the equations as a discrete model. The environment is a circular arena with a radius of $R_{arena}$ and without any obstacles in it. Neither the pursuers nor the target can get out of the arena: if they take an action that would end up outside $R_{arena}$, their position is updated to be on the nearest arena border.





We assume that the pursuers can differentiate the agents from the target. We further assume that each pursuer is equipped with a sensor that provides the relative position of the target as well as every other pursuer. All sensors provide ground truth data without noise, unaffected by occlusions. The state representation of each agent is detailed in Sec.~\ref{subsec:state_representation}.



 
 


\section{Deep Reinforcement Learning for Pursuit}
\label{sec:deep_rl}

We formulate the task for a single pursuer to be a Markov Decision Process (MDP) defined by tuple $\{\mathcal{S},\mathcal{A}, R, \mathcal{P}, \gamma\}$ where $s_t \in \mathcal{S}$, $a_t \in \mathcal{A}$, $r_t \in R$ are state, action and reward observed at time $t$, $\mathcal{P}$ is an unknown transition probability from $s_t$ to $s_{t+1}$ taking $a_t$, and $\gamma$ is a discount factor. The DRL goal is to maximise the sum of future rewards $R = \sum_{t=0}^{T}\gamma^tr_t$, where $r_t$ is provided by the environment at time $t$. Actions are sampled from a deep neural network policy $a_t\sim\pi_\theta(s_t)$, where $a_t$ is the angular velocity $\omega$ of an individual pursuer, which is saturated to be in the interval $[\omega_{min}, \omega_{max}]$. 

\subsection{Multi-Agent Deep Reinforcement Learning}

As the Deep RL algorithm, we use Twin Delayed Deep Deterministic Policy Gradient approach (TD3)\cite{fujimoto2018addressing}, which is an improvement over DDPG~\cite{lillicrap2015continuous}, designed to reduce the overestimation of the value function. We consider all agents to be homogeneous which allows us to use shared experience to train all agents. This allows the agents to train faster, as well as gathering more information from every step in the environment. All agents are governed with the same policy, however, at each time step the agents use their local observations to individually take actions, resulting in a decentralized system.
For each number of agents we train a different policy, which results in a total of $n_{max}$ policies, where $n_{max}$ is the maximum number of pursuers we analyze in this paper. This was because the length of the state representation changes based on the number of pursuers in the game. 

\subsection{State Representation}
\label{subsec:state_representation}

\begin{figure}[ht!]
\centering
\includegraphics[trim=0 0 0cm 0cm, clip, scale=0.45]{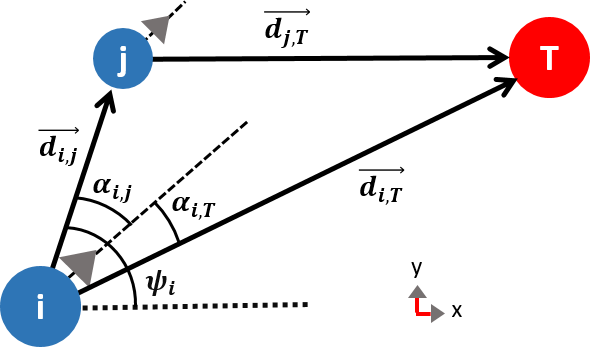}
\caption{The state space for each agent $i$. $T$ denotes the target and $j$ denotes another agent}
\label{fig:engagement}
\end{figure}

The state of a pursuer $i$, assuming a total of $n$ pursuers, is given by $s_i=[\psi_i, \dot{\psi}_i, s_{i,T}, s_{i,1}, s_{i,2}, .. , s_{i,{n-1}}]$, where $\psi$ is the heading with respect to a fixed world frame, $s_{i,T}$ is the state of the target relative to pursuer $i$ and $s_{i,j}$ is the state of pursuer $j$ relative to pursuer $i$. (Time indices are dropped for the sake of clarity). The relative state of the target with respect to pursuer $i$ is $s_{i,T}=[d_{i,j}, \dot{d}_{i,T},\alpha_{i,T}, \dot{\alpha}_{i,T}]$ and the relative state of pursuer $j$ with respect to pursuer $i$ ($i \neq j$) is $s_{i,j}=[d_{i,j}, \alpha_{i,j}]$, where $d_{i,j}$ is the Euclidean distance between pursuers $i$ and $j$ and $\alpha_{i,j}$ is the heading error defined as the angle between the heading of pursuer $i$, and the vector between $i$ and $j$, as shown in in Fig~\ref{fig:engagement}. The state representation consists of a total of $2n+4$ variables, which scales linearly with the number of pursuers $n$. 

An important consideration is how the $s_{i,j}$ are ordered in the state representation $s_i$. A straightforward way would be to assign a unique identifier to each pursuer and always represent them in the same order. However, this leads to inefficiencies in learning as neural networks typically are not permutation-invariant when operating on sets~\cite{zaheer2017deep}. To illustrate this, consider swapping the poses of two pursuers with everything else being the same. With unique identification ordering, the resulting state would be different than the original, whereas since we have homogeneous agents, the state should not change. 
To tackle this problem we assign $j$ values for each observation by sorting each other pursuer with respect to their relative angle $\alpha$.






The observation of the pursuers was designed to be easily applicable on real platforms and is used commonly in conventional single-agent pursuit~\cite{shneydor1998missile}. The agent observations do not require localization with respect to a global frame, as local observations $d$ and $\alpha$ are not referenced in global coordinates and can be extracted using onboard sensors such as laser scanners or cameras. Recent work\cite{drone_detection} demonstrates the feasibility of acquiring measurements such as range and the relative angle between the pursuers, using only embedded sensors in drones. For the heading $\psi$, a directional sensor would be needed, such as a magnetometer.


\subsection{Reward Structure}
\label{subsec:reward}
At each time step, each agent individually receives a reward designed to incentivize the capture of the evader and encourage a good formation of pursuers. The reward function is:
\[
    r_i= 
\begin{cases}
    r_{captor}\text{,}& \text{if } d_{i,T} \leq d_{cap}\\
    r_{helper}\text{,}& \text{if } d_{j,T}
    \leq d_{cap} \text{,} \, \exists j \neq i\\
    -w_q \, q - w_d \, d_{i,target}\text{,}& \text{otherwise}
\end{cases}
\]
At each step that the target is not captured, each and every agent receives a negative reward that is a weighted linear combination of an individual reward (its distance to the target $d_{i,target}$) and a group reward (q-score~\cite{kamimura_ohira_2019}, which we call the formation score in our work). The formation score is a scalar number in the range $[0,2]$, which provides a metric for evaluating the fitness of a formation of the pursuers (lower is better). The formation score ($q$) is defined as:
\begin{equation}
    q = \frac{1}{n} \sum^{n}_{i=1} (\uvec{$d$}_{\textbf{0T}} \cdot \uvec{$d$}_{\textbf{iT}} + 1)
\end{equation}
where $\uvec{$d$}_{\textbf{iT}}$ denotes a unit vector pointing in the direction from agent $i$ and the target, and $n$ is the number of agents. In this equation, the closest agent to the target is defined as agent $0$. The formation score encourages agents to spread around the target (i.e., approach the target from different directions) and penalizes the angular proximity between agents. We introduce the formation score when there are at least two agents, as the formation score for one agent is not defined. Early experiments with the formation score showed that when the formation is the only component of the reward, pursuers would only form a good formation but would not make an attempt to capture the evader. To avoid this situation, we penalize the distance to the target, which helps encourage the agents to get close to the evader while being in a good formation. The weights $w_q$ and $w_d$ were chosen such that when the agents are close to the target, the reward is dominated by the formation score, encouraging good formation. However, when the agents are far from the target, the reward is dominated by the distance to the target, encouraging the agents to move closer to the evader. We analyze the effect of the formation score on pursuit performance in Sec.~\ref{subsec:results_q_score}.

If the target is captured at a time step, then the pursuer who captures the evader receives the reward $r_{captor}$, while the rest of the agents receive $r_{helper}$, such that $r_{captor} > r_{helper}$. 
This encourages each pursuer to go for the final capture while encouraging collaboration.


\subsection{Curriculum Learning}
\label{subsec:curriculum_learning}

We apply a curriculum for learning by starting from an easier version of the task and gradually increasing the difficulty until the actual difficulty is achieved. There are two main factors in determining the difficulty of a pursuit-evasion game: the relative speed of the target with respect to the pursuers and the capture radius $d_{cap}$. We vary the capture radius by starting from a large radius (so it is easier to capture the target), then gradually making it smaller. This encourages agents to not adopt a straightforward chasing tactic at the beginning of learning but to form more sophisticated behaviors, which could be transferred to smaller capture radii. We also experimented with reducing the pursuer speed to reduce the difficulty of the task, however, we found that pursuers mostly learned to follow the evader directly, and it was harder to explore more sophisticated behaviors afterward.

Curriculum learning helps exploration, especially during the early stages of learning, because early, on it helps the pursuers to capture the target, which would take a longer time in the actual, and more difficult scenario. This helps alleviate the sparse reward problem, which is a well-known challenge in DRL~\cite{nair2018overcoming}. We analyze the effect of curriculum learning on the pursuit performance in Sec.~\ref{subsec:results_curriculum}. 




\section{Simulation Experiments}
\label{sec:experiments}

We use the following simulation parameters: $T_{timeout}=500$ iterations, rewards $r_{captor}$, $r_{helper}$; the weights $w_q$ and $w_d$ were set to 10, 100, 0.1, 0.002 respectively. The capture radius $R_{cap}$ was set to 30 pixels for testing, although this value was varied as part of curriculum learning. For fixed linear velocity cases, the speed of the pursuers $v_p$ was 10 pixels per timestep, while the target's speed $v_T$ varied from 0 to 20 pixels per timestep. The maximum angular rate $\omega_{max}$ was fixed at $\pi/10$ per timestep. The number of pursuers $n$ varied between 1 and 8, initialized at random positions within a circular area with a radius of 100 pixels, while the evader was initialized at a random position between the arena boundary and an inner circle with a radius of 300 pixels. Arena radius $R_{arena}=430$ pixels, except for the results in Sec \ref{subsec:arena_size}. We assume each pursuer can observe all other pursuers, except for in Sec. \ref{subsec:scalability}.

\subsection{Evader Behaviors}
\label{subsec:evader_behavior}

We implemented two behavior modes for the evader: Fixed Paths and Repulsive. For both, we vary the relative speed of the evader from $0.8$ to $2$ times the pursuer speed, with a step size of $0.2$. We conduct 100 trials for every speed level.

\begin{figure}[ht!]
\centering
\includegraphics[ scale=0.25]{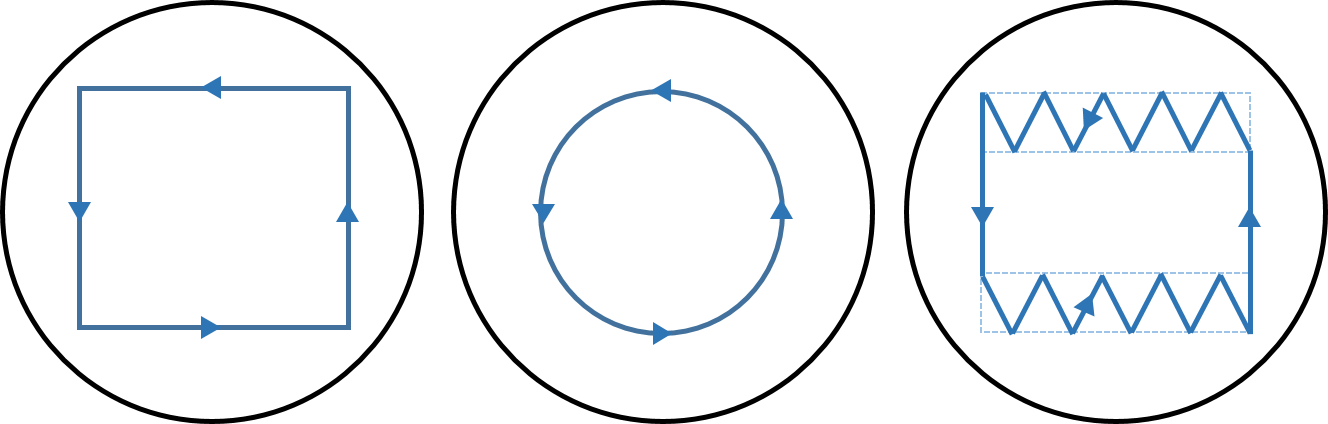}
\caption{Paths used by the evader for the fixed paths benchmark.}
\vspace{-0.1cm}
\label{fig:fixed_paths}
\end{figure}

\textbf{Fixed Paths}: We propose a benchmark where the evader follows three predefined paths, as shown in Fig.~\ref{fig:fixed_paths}.

\textbf{Repulsive}: We use a potential field method with repulsive forces only to find a motion vector. Each pursuer exerts a repulsive force in the direction of the vector between the pursuer and the evader. The arena boundary also exerts a force so that the evader can avoid the wall. These forces decrease proportionally to the distance squared. The resultant vector is calculated by:
\begin{equation}
    \vec{v} = \sum_j (\frac{\vec{a}_{j} - \vec{e}}{d_j^2}) + \frac{\uvec{$\gamma$}R_{arena}  - \vec{e}}{d_{w}^2}
\end{equation}
where $\vec{e}$ is the current position of the evader, $\vec{a}_{j}$ is the position of agent $j$ and $d_j$ is the distance to $\textit{a}_j$. $\gamma$ is the direction of the agent to the closest point on the wall, $\uvec{$\gamma$}$ is the unit vector rotated by an angle of $\gamma$ and $d_w$ represents the distance of the agent to the wall.

\subsection{Baseline methods}
\label{sec:baseline}

We implemented three baseline methods: Two classical (Janosov \cite{Janosov2017} and Angelani \cite{angelani2012collective}) and a DRL (Hüttenrauch \cite{Huttenrauch2019}) method. The approach by Hüttenrauch~\cite{Huttenrauch2019} was trained on our simulation environment using their \textit{communication} set. This observation vector included relative angle, the distance towards the target, and the heading of each of the pursuers. It should be noted that this observation set included more information than our model. Furthermore, this information (orientation of neighbors) would likely require explicit communication in a real-world application since it can not be easily estimated from current embedded sensors. We trained the policy for 4 million timesteps (same as our approach, around 4 times as long as their original paper) without curriculum learning (no curriculum learning was used in their original work) and presented the best simulation results.

We adapted the classical methods to use a non-holonomic model. As these methods are designed for omnidirectional agents, they are not directly comparable to our solution. Therefore, we convert the outputs of these models into the unicycle model using the following equations:
\begin{equation}
    \psi_{desired} = \arctan{\frac{dy}{dx}}
\end{equation}
\begin{equation}
    \omega = K * (\psi - \psi_{desired})
\end{equation}
The omnidirectional models have two outputs, $dx$ and $dy$, which are the velocity in the $x$ and $y$ direction respectively. From this, we find the desired heading $\psi_{desired}$ of the omnidirectional controller. We use a P controller on the error between the desired heading $\psi_{desired}$ and current heading $\psi$. We tune gain $K$ such that the number of captures is maximized in simulation trials.

\section{Simulation Results}
\label{sec:results}

In this section we evaluate our approach against baseline algorithms presented in Sec.\ref{sec:baseline}. In Sec.\ref{fixed} we analyze the performance on a fixed paths benchmark, followed by analysis with a repulsive evader model. We conduct the following analyses on the capture performance: effect of number of pursuers (Sec.\ref{subsec:number_of_agents}), arena size (Sec.\ref{subsec:arena_size}), relative evader speed (Sec.\ref{subsec:relative_evader_speed}), scaling number of agents without retraining (Sec. \ref{subsec:scalability}), use of curriculum learning (Sec.\ref{subsec:results_curriculum}), and use of formation score (Sec.\ref{subsec:results_q_score}). Finally, we qualitatively describe the emergent behaviour of the multi-agent system in Sec.\ref{subsec:qualitative}.

\subsection{Fixed Paths}
\label{fixed}

\begin{figure}[h!]
\centering
   \begin{subfigure}{0.49\linewidth}
   \centering
   \includegraphics[trim=0.48cm 0 0.3cm 0.2cm, clip, width=\linewidth]{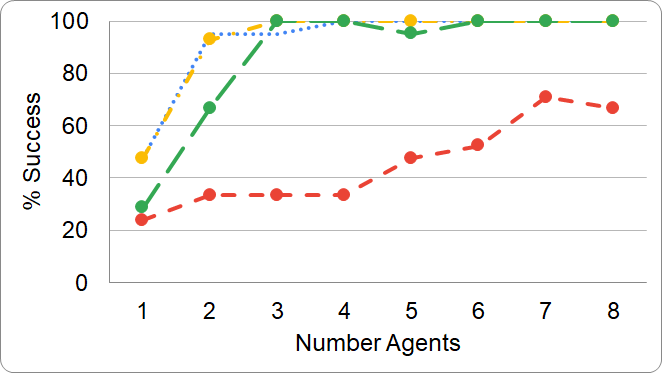}
\end{subfigure}
\begin{subfigure}{0.49\linewidth}
   \centering
   \includegraphics[trim=0.5cm 0 0.3cm 0.2cm, clip,width=\linewidth]{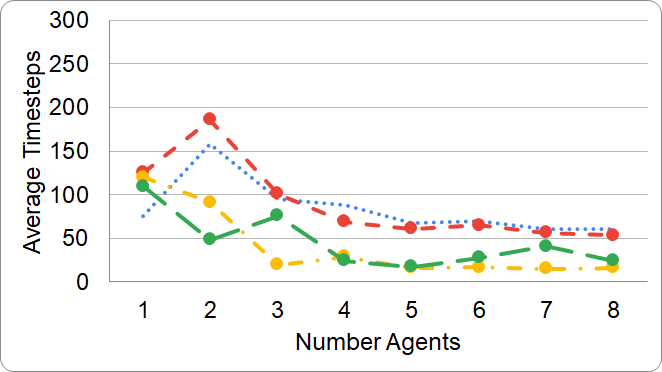}
\end{subfigure}
\begin{subfigure}{1.0\linewidth}
   \centering
   \includegraphics[width=0.7\linewidth]{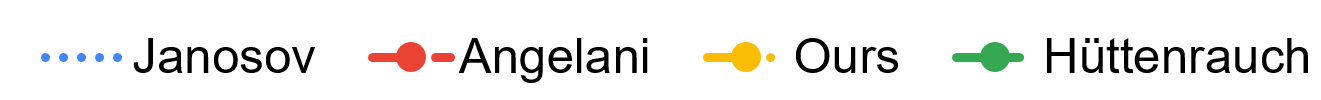}
\end{subfigure}
\caption{Success rate and the average timesteps using the fixed paths benchmark, for different number of agents}
\centering
\label{fig:fixedtraj}
\end{figure}

In these experiments, the evader followed fixed paths, as explained in Sec.\ref{subsec:evader_behavior}. The pursuers were trained on the repulsive evader only but were tested on the fixed paths benchmark. The capture rate and the average number of steps with respect to the number of pursuers are shown in Fig. \ref{fig:fixedtraj}. While Janosov~\cite{Janosov2017} get above $95\%$ on the fixed paths benchmark for $N > 2$, Angelani~\cite{angelani2012collective} does not perform well on this benchmark. Both our approach and Hüttenrauch~\cite{Huttenrauch2019} complete the task successfully for $n \ge 3$. The average timesteps to capture for both DRL approaches is significantly lower than the other approaches for the fixed paths benchmark, likely because the classical algorithms attempt to corral the target. In contrast, the DRL based approaches tended to be more aggressive and intercept the evader quickly along the fixed paths.


\subsection{Effect of the Number of Pursuers}
\label{subsec:number_of_agents}

\begin{figure}[H]
\centering
   \begin{subfigure}{0.49\linewidth}
   \centering
   \includegraphics[trim=0.48cm 0 0.3cm 0.2cm, clip, width=\linewidth]{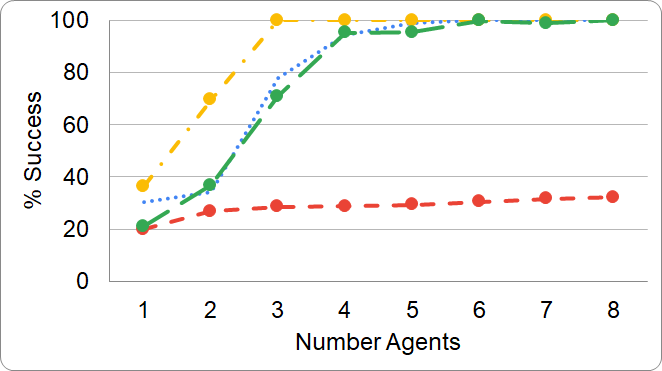}
\end{subfigure}
\begin{subfigure}{0.49\linewidth}
   \centering
   \includegraphics[trim=0.48cm 0 0.3cm 0.2cm, clip, width=\linewidth]{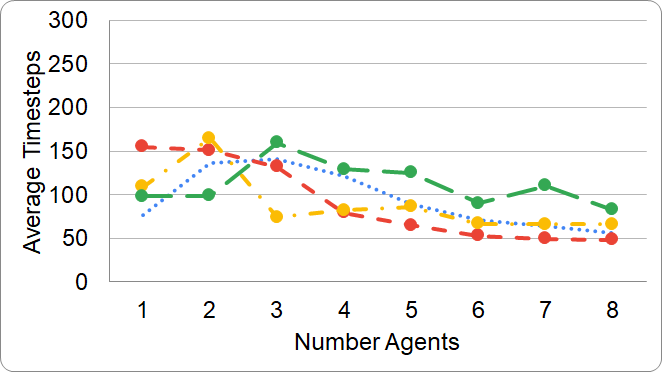}
\end{subfigure}
\begin{subfigure}[H]{1.0\linewidth}
   \centering
   \includegraphics[width=0.7\linewidth]{Updated Figures/Legend_Updated.png}
\end{subfigure}
\caption{Success rate and the average timesteps taken for the repulsive evader, for different number of agents.}
\centering
\label{fig:number_of_agents}
\end{figure}

The success rate and average timesteps to capture the repulsive evader with respect to the number of pursuers are shown in Fig.~\ref{fig:number_of_agents}. Our approach outperforms the competing approaches in cases with a lower number of agents in terms of the success rate. Hüttenrauch~\cite{Huttenrauch2019} performs well with a larger number of agents, however, struggles with fewer agents. Janosov~\cite{Janosov2017} completes the task with a success rate above $94\%$ for $n\ge4$. However, it does not perform well with fewer than four agents. Angelani~\cite{angelani2012collective} does not perform well in this benchmark. Furthermore, with one or two agents, all methods showed poor performance, as it is difficult to chase a faster evader with few agents. Our approach takes more time on average to complete the capture compared to~\cite{Janosov2017}. 

%
\subsection{Effect of arena size}
\label{subsec:arena_size}

\begin{figure}[H]
\centering
   \begin{subfigure}{0.49\linewidth}
   \centering
   \includegraphics[trim=0.48cm 0 0.3cm 0.2cm, clip,width=\linewidth]{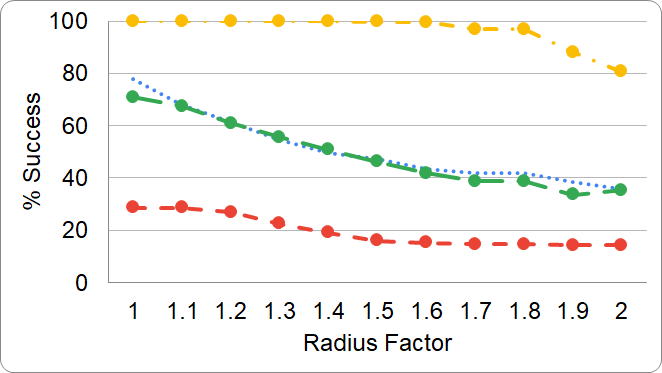}
\end{subfigure}
\begin{subfigure}{0.49\linewidth}
   \centering
   \includegraphics[trim=0.48cm 0 0.3cm 0.2cm, clip,width=\linewidth]{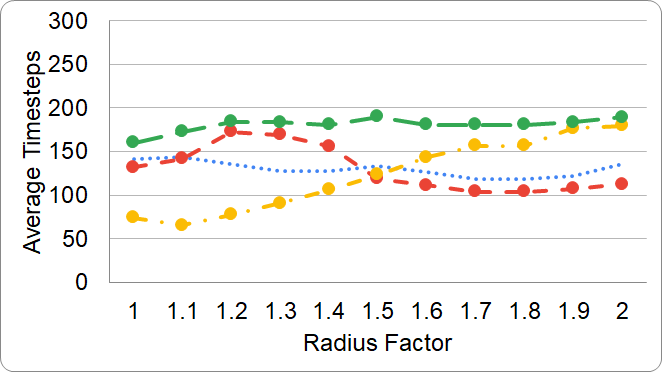}
\end{subfigure}
\begin{subfigure}[H]{1.0\linewidth}
   \centering
   \includegraphics[width=0.7\linewidth]{Updated Figures/Legend_Updated.png}
\end{subfigure}
\caption{Success rate and average timesteps to capture with respect to the multiplicative radius factor over $R_{arena}$. Experiments are conducted with 8 pursuers and the repulsive evader model.
}

\centering
\label{fig:radius}
\end{figure}

As we showed in Sec. VI.B, with an increasing number of agents, the task decreases in difficulty for a fixed arena size. Therefore, with a larger number of agents, most approaches can perform the task successfully. The pursuit-evasion game is played in an obstacle-free arena, but the agents can use the arena boundaries to constrain the evader movements. In this section, we investigate the effect of larger arena sizes using $n=3$ agents. We do not retrain our agents on the new arena size but use the model trained on the original radius size $r_{area}$. As shown in Fig. \ref{fig:radius}, our approach comfortably outperforms the other approaches in terms of success rate as the arena size increases. This shows that the learned policy can generalize to larger arenas. However, the average number of timesteps to capture for our approach is consistently higher than other approaches. We attribute this result to using only the successful captures to obtain the average number of timesteps: our approach can likely find solutions to more difficult problems at the expense of increased average duration to secure the capture.




\subsection{Effect of Relative Evader Speed}
\label{subsec:relative_evader_speed}


\begin{figure}[h!]
\centering
   \begin{subfigure}{0.49\linewidth}
   \centering
   \includegraphics[trim=0.48cm 0 0.3cm 0.2cm, clip, width=\linewidth]{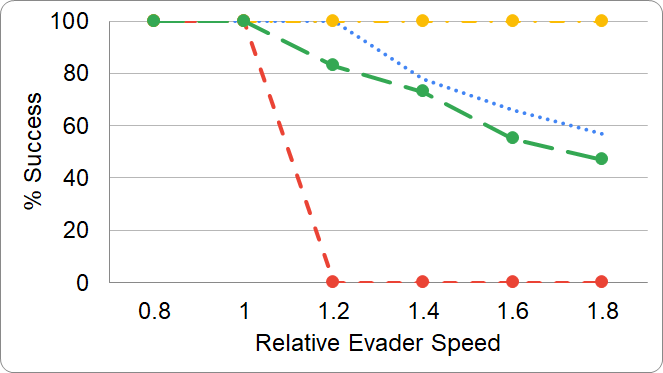}
\end{subfigure}
\begin{subfigure}{0.49\linewidth}
   \centering
   \includegraphics[trim=0.48cm 0 0.3cm 0.2cm, clip, width=\linewidth]{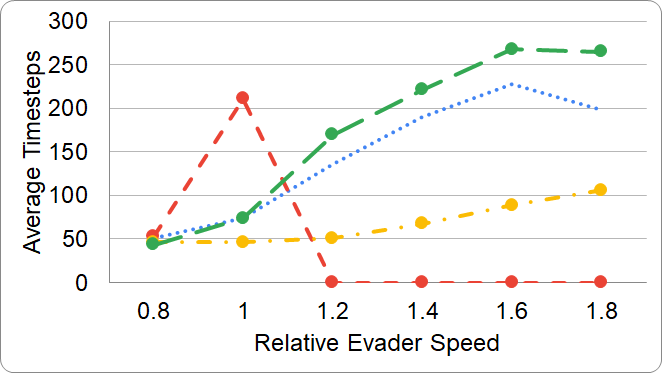}
\end{subfigure}
\begin{subfigure}{1.0\linewidth}
   \centering
   \includegraphics[width=0.7\linewidth]{Updated Figures/Legend_Updated.png}
\end{subfigure}
\caption{Success rate and the average number of timesteps to capture is shown with respect to the ratio of evader speed to pursuers' speed. Experiments are run with three agents. Our approach achieves 100\% accuracy for all cases in this analysis.}
\centering
\label{fig:relative_speed}
\end{figure}

In this section, we examine at the effect of relative evader speed on capture success, for $n=3$ pursuers. As shown in Fig.~\ref{fig:relative_speed}, while our approach had 100\% success rate at all speed levels, all other methods had a drop-off at faster evader speeds. As expected, our approach was the fastest in capturing the target (only considering successful episodes).

\subsection{Scaling number of agents without retraining}

\label{subsec:scalability}

\begin{figure}[H]
\centering
   \begin{subfigure}{0.49\linewidth}
   \centering
   \includegraphics[trim=0.48cm 0 0.3cm 0.2cm, clip, width=\linewidth]{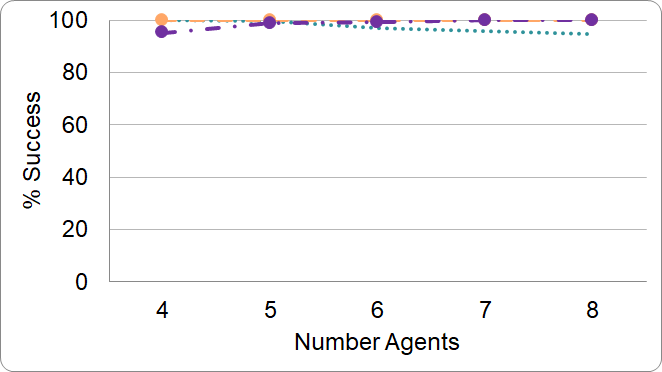}
\end{subfigure}
\begin{subfigure}{0.49\linewidth}
   \centering
   \includegraphics[trim=0.48cm 0 0.3cm 0.2cm, clip, width=\linewidth]{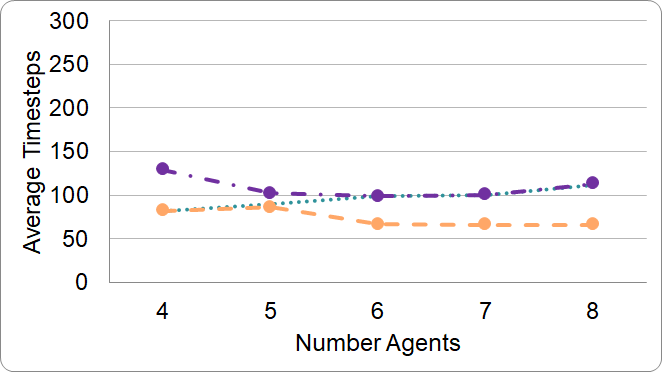}
\end{subfigure}
\begin{subfigure}[H]{1.0\linewidth}
   \centering
   \includegraphics[width=\linewidth]{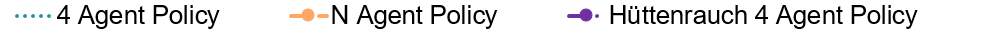}
\end{subfigure}
\caption{The success rate and average timesteps on the repulsive evader benchmark, comparing our full approach (trained on the appropriate amount of agents), the approach by Hüttenrauch\cite{Huttenrauch2019} which was trained on four agents and our approach which was limited to only see the closest four agents during execution.}
\centering
\label{fig:scalable}
\end{figure}

In this section, we examine the scalability of our approach to an environment with more agents than the network was trained on. For the 4 agent policy in Fig. \ref{fig:scalable}, the agents are trained in the 4 agent setting. At test time, the number of agents increases (as indicated on the horizontal axis), but each agent can only observe their 4 closest neighbors. In contrast, both the $N$ agent policy and Hüttenrauch~\cite{Huttenrauch2019} were given information about all pursuers. 
The success of our approach decreases with a larger number of agents, by around 5\% when there are more agents, as the agents cannot coordinate fully. The other approaches reach 100\% for a larger number of agents. 
The four agent policy requires a larger average number of timesteps to capture the target compared to a more specialized policy, and this performance penalty increases as more agents appear in the environment compared to the number of agents during training. However, the 4-agent policy tends to perform similarly to \cite{Huttenrauch2019}, which was designed for scalability.

\subsection{Effect of Curriculum Learning}
\label{subsec:results_curriculum}

\begin{figure}[h!]
\centering
\includegraphics[trim=0.48cm 0.5cm 0.5cm 2.5cm, clip, width=0.85\linewidth]{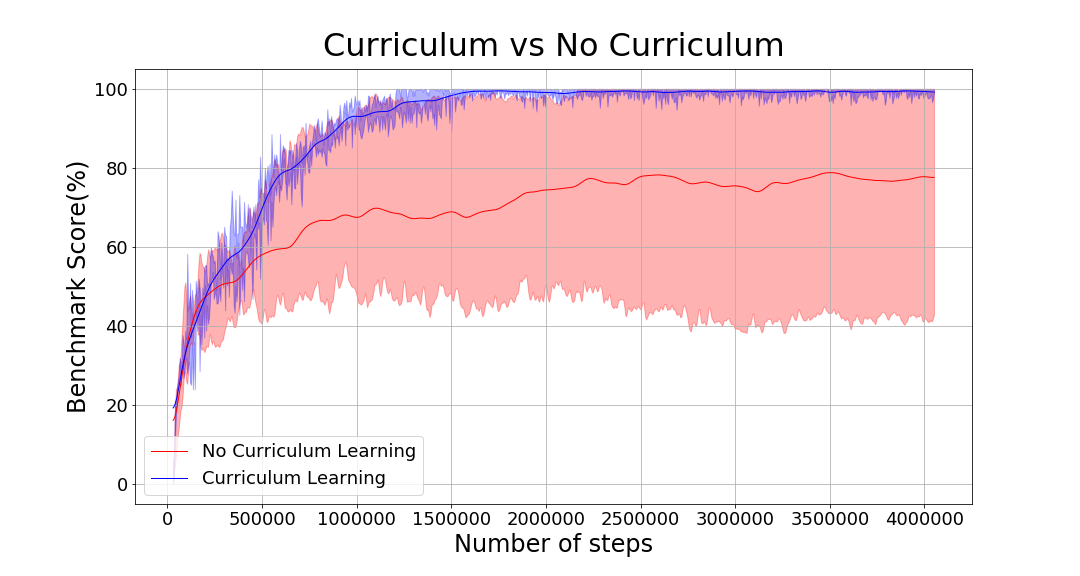}
\caption{Comparison of capture success rate with and without curriculum learning, with respect to the number of training steps. With curriculum learning, the benchmark scores are much higher and more consistent.}
\label{fig:Curr_Comparison}
\end{figure}

Fig.~\ref{fig:Curr_Comparison} compares the effect of using our curriculum learning strategy described in Sec~\ref{subsec:curriculum_learning}, for $n=3$ agents. The network was trained 3 times. At regular intervals, we stop training and evaluate the policy on the repulsive evader benchmark. The results show that curriculum learning is beneficial for capture performance: it converges to about 100\% success rate after 1.5 million training steps, whereas without curriculum learning, the average success rate was below 80\% even at 4 million training steps. Furthermore, the performance with curriculum learning was much more consistent, as evidenced by the low variance among the three runs. 

\subsection{Effect of Formation Score in Reward Function}
\label{subsec:results_q_score}

\begin{figure}[h!]
\centering
   \begin{subfigure}{0.49\linewidth}
   \centering
   \includegraphics[trim=3cm 3cm 4.5cm 4.5cm, clip,width=\linewidth]{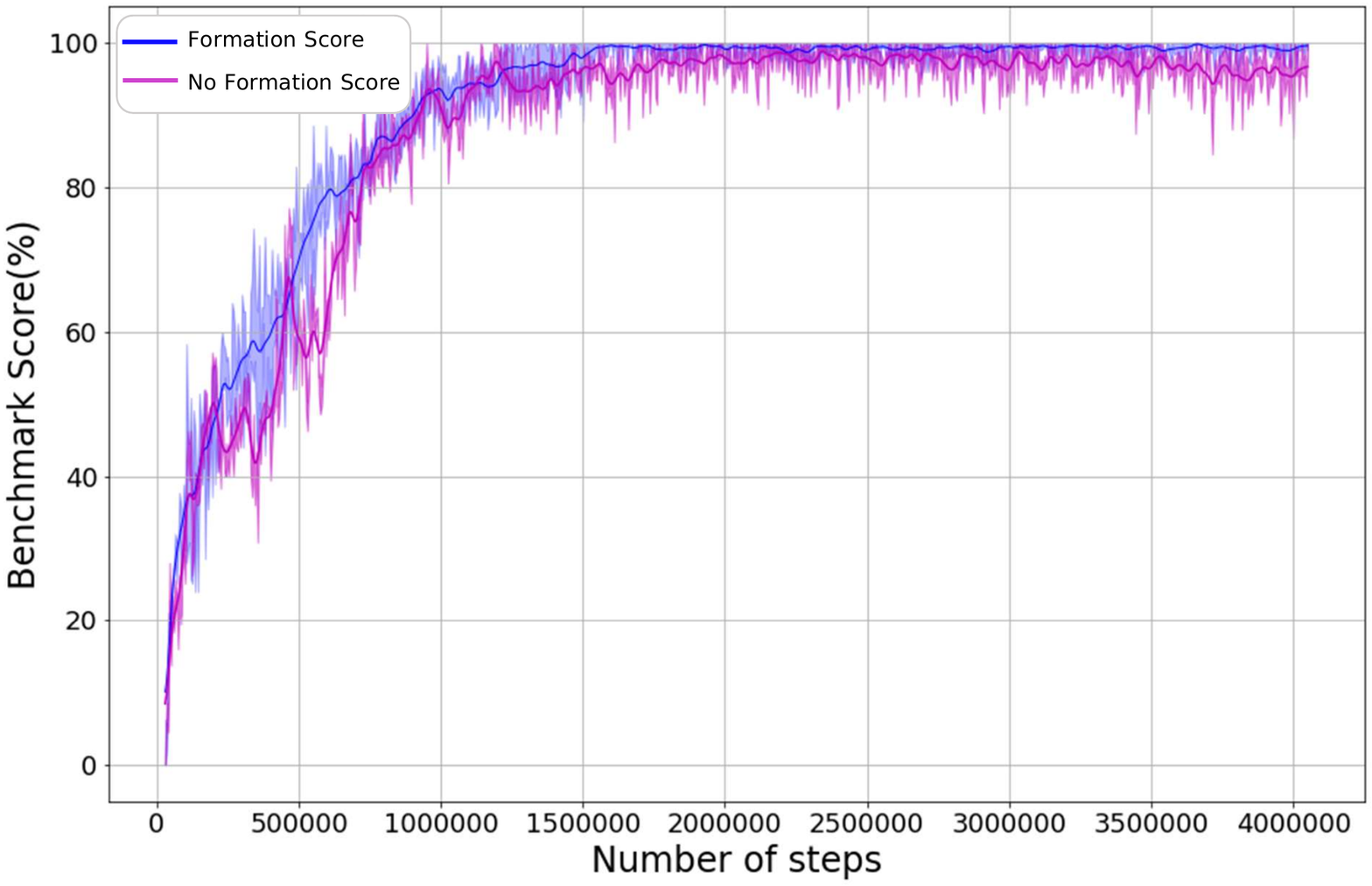}
\end{subfigure}
\begin{subfigure}{0.49\linewidth}
   \centering
   \includegraphics[trim=6.5cm 6.5cm 9.3cm 5.5cm, clip,width=\linewidth]{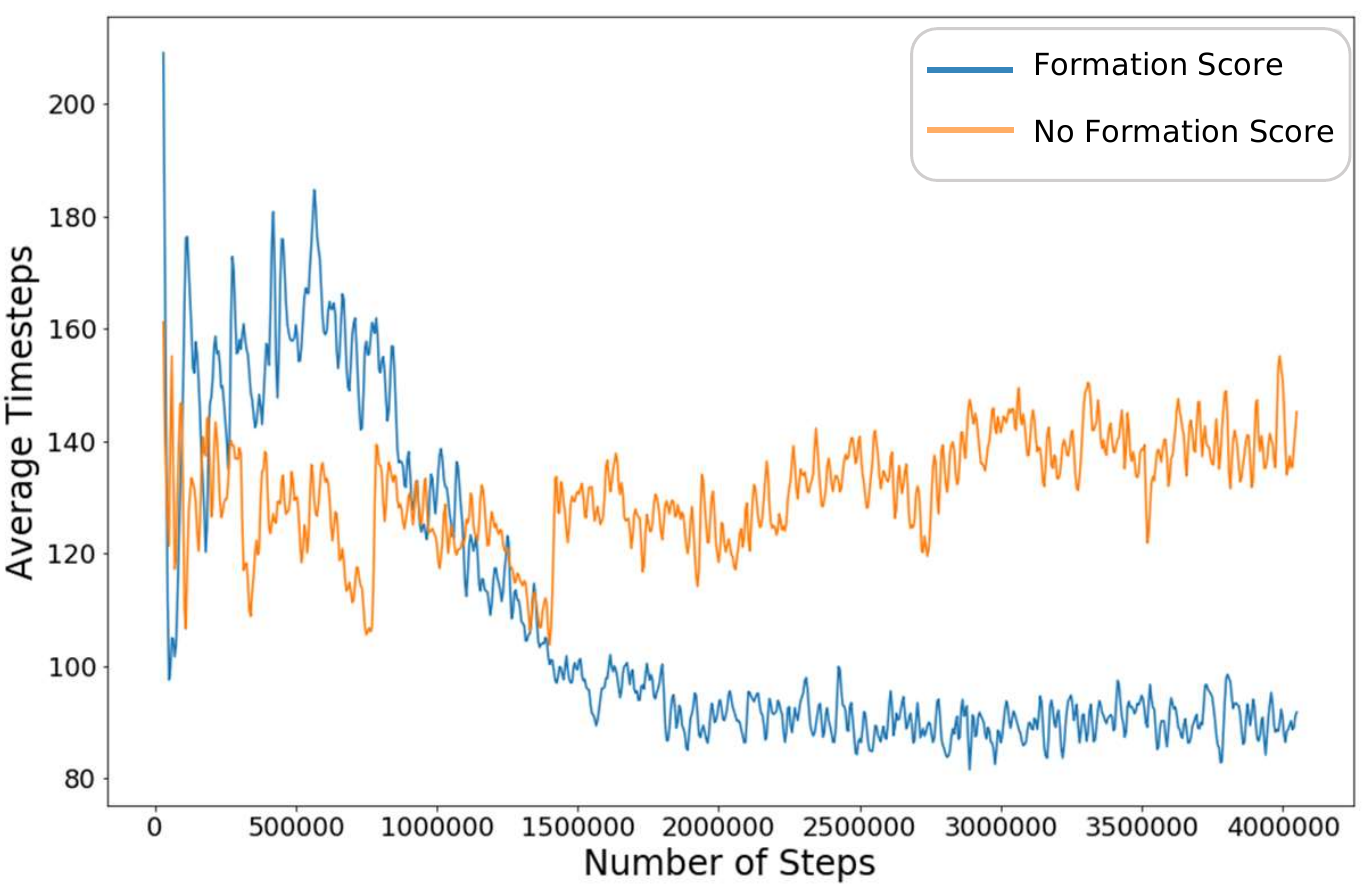}
\end{subfigure}
\caption{Using a formation score as a dense reward results in more captures, in less number of timesteps on average.}
\centering
\label{fig:formation_score_comparison}
\end{figure}

As described in Sec.~\ref{subsec:reward}, we provide a partial reward at every timestep in order to encourage good formations. We analyze the effect of supplying this dense reward component to each agent. Fig.~\ref{fig:formation_score_comparison} compares the evolution of the capture performance with and without the formation score with respect to the number of training steps. These experiments were conducted with $n=3$ agents. As shown in the Fig. \ref{fig:formation_score_comparison}, benchmark scores were slightly higher when the formation score is used as part of the reward. Furthermore, when the formation score is used, the average capture time for successful episodes is decreased.

\subsection{Qualitative Analysis of Emergent Behavior}
\label{subsec:qualitative}

We observe two interesting learned emergent behaviors that often lead to successful captures: ambushing and splitting up.

\begin{figure}
\centering
\includegraphics[trim=0.3cm 0.25cm 0.55cm 0.35cm, clip, width=0.32\linewidth]{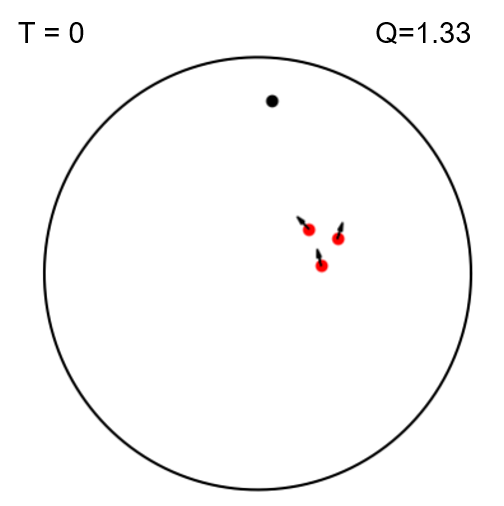}
\includegraphics[trim=0.3cm 0.15cm 0.55cm 0.35cm, clip, width=0.32\linewidth]{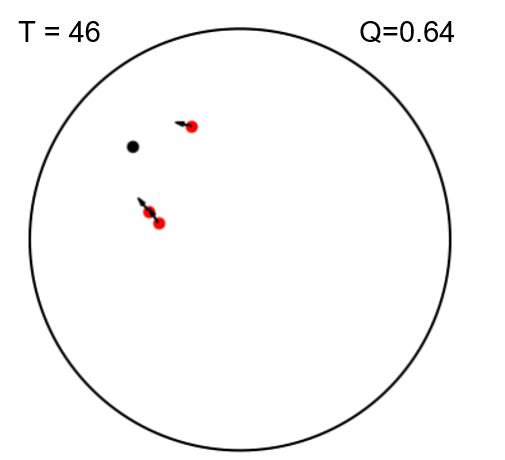}
\includegraphics[trim=0.25cm 0.25cm 0.55cm 0.35cm, clip, width=0.32\linewidth]{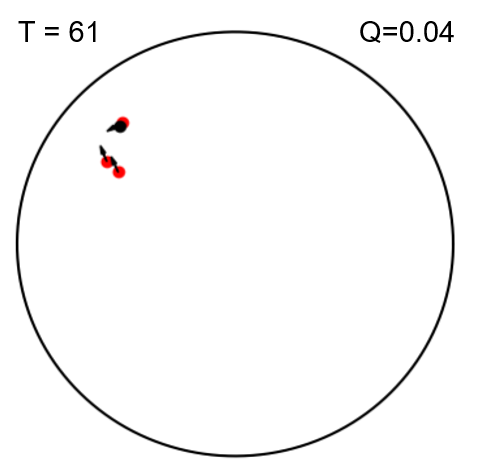}\par
\caption{``Split Up" strategy learned by three pursuers. Timestep (T) and formation scores (Q) are shown at three snapshots. The target is shown as the black circle. The agents start in a random direction (Left), push the agent towards the wall splitting into two groups (Middle) before going for the capture (Right).}
\label{fig:3agents}
\end{figure}

Fig.\ref{fig:3agents} shows the splitting up behavior with $n=3$ agents. This behavior was more common with a smaller number of agents. The agents tend to split up into two groups, trying to push the evader into a wall before attempting to block the two opposite directions. This tactic works well as the evader is backed against the wall and has limited room to escape.

\begin{figure}
\centering
\includegraphics[trim=0.3cm 0.5cm 0.55cm 0.35cm, clip,  width=0.32\linewidth]{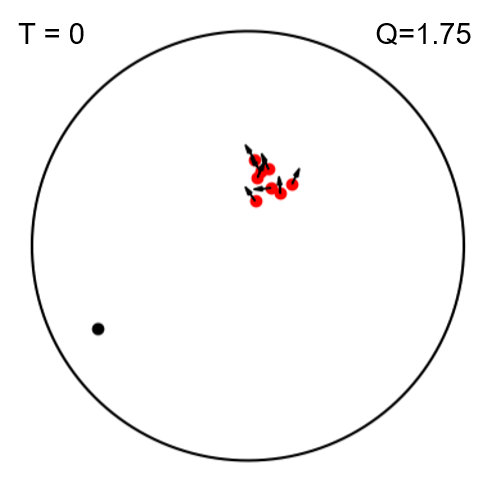}
\includegraphics[trim=0.3cm 0.15cm 0.5cm 0.35cm, clip,width=0.32\linewidth]{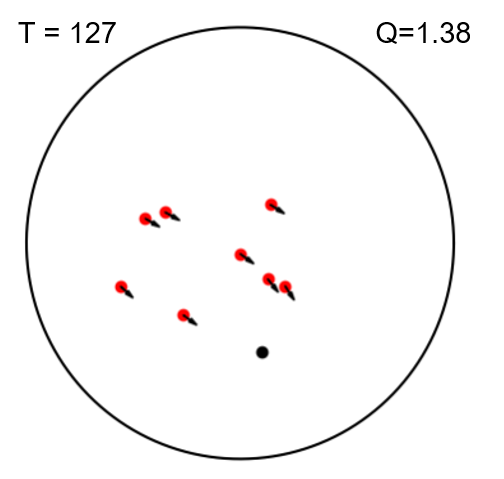}
\includegraphics[trim=0.3cm 0.3cm 0.45cm 0.35cm, clip,width=0.32\linewidth]{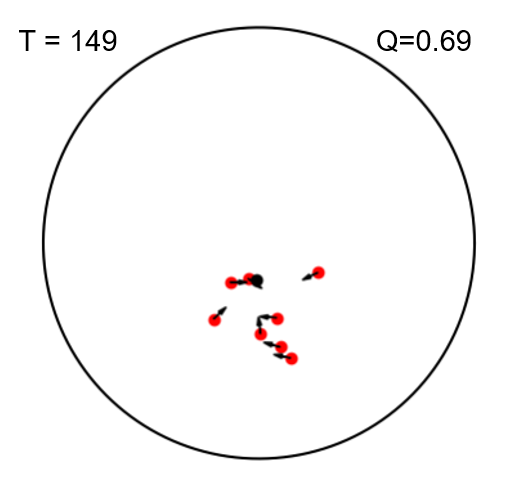}\par
\caption{``Ambush" strategy learned by 8 pursuers. Timestep (T) and formation scores (Q) are shown at three snapshots. The target is shown as the black circle. The agents start in random directions (Left), move as a circle (Middle) ambush the target and capturing it (Right)}
\label{fig:8agents}
\end{figure}

Fig. \ref{fig:8agents} shows the ambushing behavior with $n=8$ agents. This behavior was more common with a larger number of agents. The agents tend to form a circle, attempting to move such that they can surround the evader and then approach from all directions. This seems to be a distinct behavior from the 'Split Up' behavior, as the agents do not use the walls as much, preferring to surround the evader as a pack, similar to pack behaviors observed in Muro's work~\cite{muro2011wolf}. When the agents execute this strategy to trap the evader, it is often very difficult for the evader to escape.

\subsection{Variable Linear Velocity}
\label{subsec:velocity_control}

Previous sections considered constant linear speed and variable angular speed for pursuers, primarily because this is an assumption for classical algorithms. We now consider the more general case, where agents can also vary their linear velocity between 0 and $v_p$. Therefore, we train the network with two outputs: linear and angular velocity. We consider the $3$ agent scenario, in which the agents achieve 100\% capture rate with and without velocity control, in both the fixed and reactive benchmarks.

\begin{figure}[t!]
\centering
\includegraphics[trim=0.3cm 0.15cm 0.55cm 0.35cm, clip,  width=0.32\linewidth]{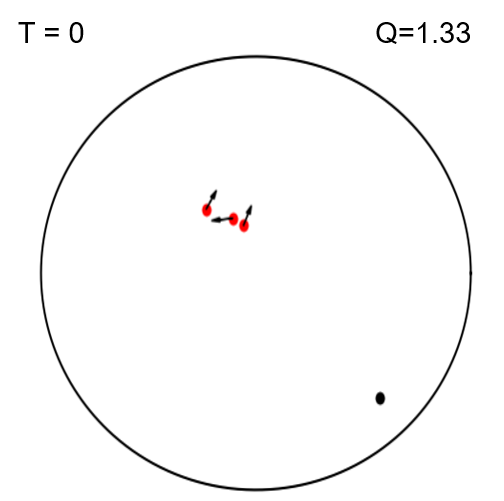}
\includegraphics[trim=0.3cm 0.3cm 0.55cm 0.35cm, clip, width=0.32\linewidth]{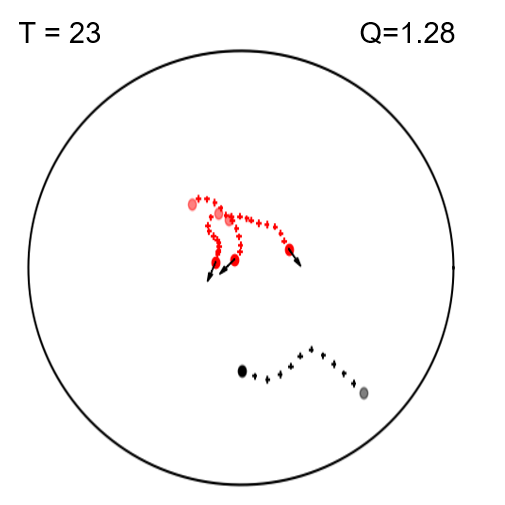}
\includegraphics[trim=0.2cm 0.2cm 0.45cm 0.3cm, clip,width=0.32\linewidth]{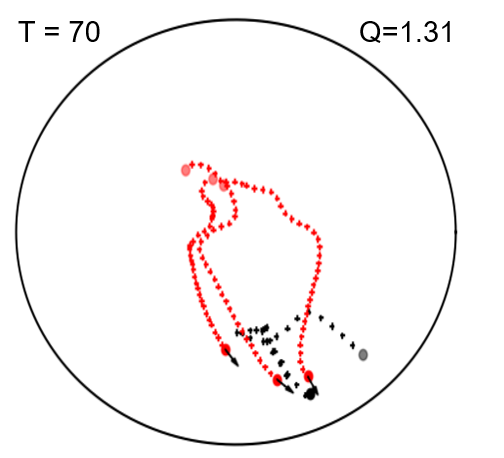}
\caption{``Stalking" strategy learned by three pursuers with velocity. Timestep (T) and formation scores (Q) are shown at three snapshots. The target is shown as the black circle. The agents start in random directions (Left), slow down and angle themselves such that they can surround the target (Middle) and capture it (Right).}
\label{fig:3velocity}
\end{figure}

The agents often displayed a ``Stalking" strategy as illustrated in Fig. \ref{fig:3velocity}. With this strategy, the agents move towards the target before slowing down and waiting until the opportunity presents itself to capture the evader. This behavior may have analogs in nature, where pursuers will stalk their prey and position themselves such to maximize the likelihood of attack\cite{Stander2004CooperativeHI}.

\section{Demonstration on Drones}
\label{sec:implementation}

We demonstrate our approach on three autonomous quadcopter drones pursuing a human-controlled target drone. Drones are typically modeled as holonomic vehicles; however, under certain conditions can act as non-holonomic vehicles (e.g. high-velocity maneuvering such as in \cite{elhennawy2017trajectory}). The use of drones, classically a holonomic vehicle, will also allow us to compare classical holonomic works (such as \cite{Janosov2017} and \cite{angelani2012collective}) to our work on the same platform in future work. Furthermore, by constraining the motion, there is an interesting property while considering a frontal camera as a sensor: the drone will move only in the direction of the field of vision, tightly coupling the perception to the movement.

We use direct sim-to-real transfer, where the policies used to control the drone behaviors are trained in the simulation environment described in Sec.\ref{sec:setup}. The input to the actor network was the normalized relative positions of the target and neighbors. The policy output for each agent is a single number, the angular velocity. We artificially constrain the motions of the pursuer drones to emulate a system with 2D agents with the unicycle kinematic model: 1) Each drone is constrained to a fixed height. The pursuer drones are at the same height, however, the evader is constrained to a different altitude, which allows the pursuers to get closer to the evader than if they were at the same altitude. 2) Angular input velocities generated by our approach are converted to input signals for low-level attitude control. 

A low-level, non-linear controller runs onboard each drone for tracking velocity reference signals: it takes the requested linear and angular velocities as input and calculates the torque and thrust for the quadcopter. Details for the low-level controller implementation can be found in~\cite{SANA10RI}. The controller is also responsible for stabilizing the quadcopter's altitude and maintaining safety. It implements collision avoidance, constrains the drones to a circular arena of $3$m radius, and limits the maximum speed to $1.2$m/s. 

The experiments took place in an indoor flight arena equipped with a motion capture system, which was used to track the pose of all agents. A centralized motion capture system was used in this implementation due to the ease of prototyping; however, all information needed by our algorithm can be captured using onboard sensors, similar to \cite{drone_detection}. Parrot AR Drone 2 was used for all drones. The behavior of each pursuer was calculated on a local computer and transmitted wirelessly to the drone at 20Hz. Details for the hardware implementation can be found in~\cite{de2018enhanced}. This setup between drones and a local computer is reminiscent of a centralized system; however, our methodology is also suitable for a decentralized system if onboard processors on each drone can be used for neural network inference. For a decentralized system, each drone would also need to be equipped with a directional sensor such as a magnetometer.

\begin{figure}[h!]
	\includegraphics[scale=0.3]{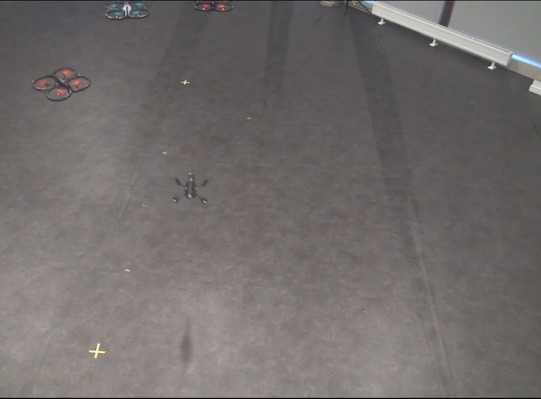}\hspace{0.01cm}
	\includegraphics[scale=0.3]{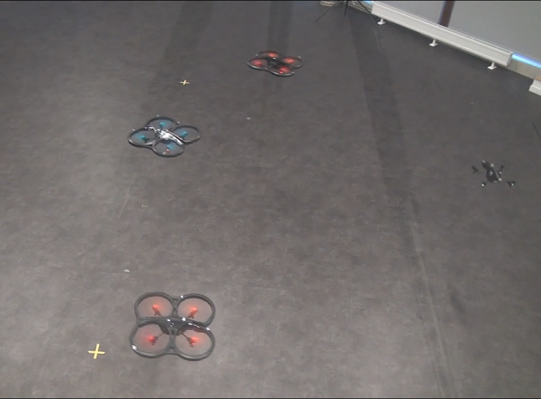}\hspace{0.01cm}
	\includegraphics[scale=0.3]{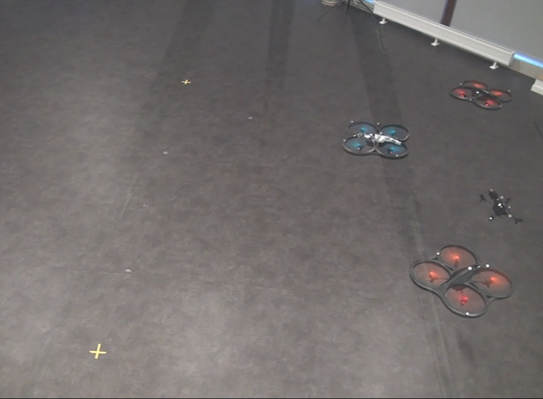}\\
	\includegraphics[trim=-1.8cm 0 0 0, scale=0.26]{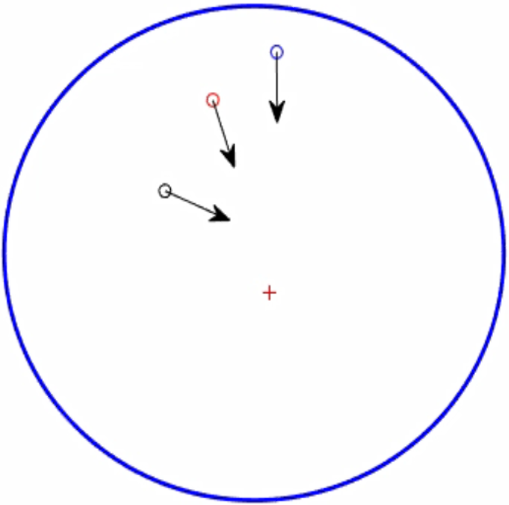}\hspace{0.3cm}
	\includegraphics[scale=0.26]{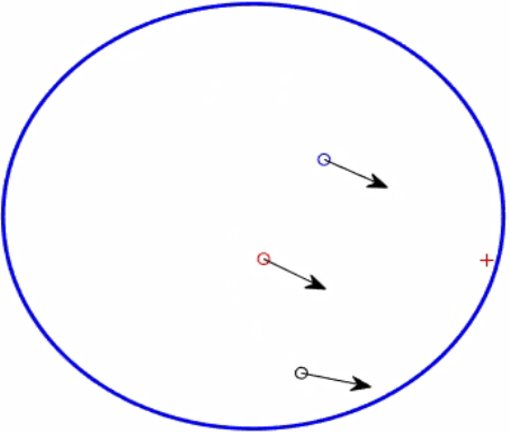}\hspace{0.5cm}
	\includegraphics[scale=0.26]{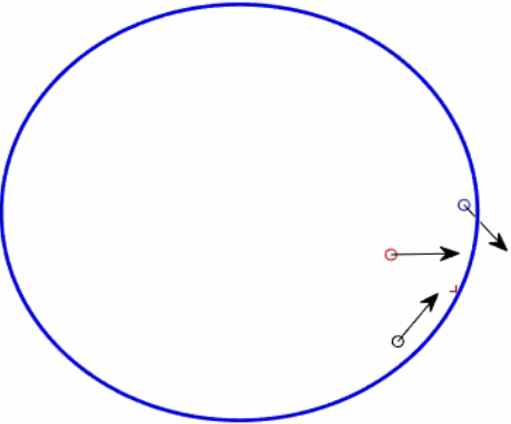}
	\caption{Snapshots from real-world demonstration with 3 motion-constrained drones. We can see the emergence of the ``Split Up" behavior: The pursuers are initially close to each other, then spread toward the target, and finally, regroup by cornering the target.}
	\label{fig:real_drones}
\end{figure}

Snapshots from a successful demonstration can be seen in Fig.\ref{fig:real_drones}. During the training of the networks, we do not consider the non-linear dynamics of the quadcopter. Direct sim-to-real transfer is possible because DRL policy provides high-level navigation decisions, while a lower-level controller manages the attitude of the drone and assures safe navigation. 

\section{Conclusion}
\label{sec:conclusion}

We proposed a DRL approach to multi-agent pursuit with non-holonomic pursuers and an omnidirectional target. We consider a decentralized system where each agent individually decides on its own actions using local observations only. Simulation experiments show that our approach, applied to non-holonomic agents, outperforms the state-of-the-art in heuristic multi-agent pursuit methods and a recent DRL based approach. Our results show that multi-agent pursuit benefits from curriculum learning and a reward based on agent formation, which we borrowed from the group-pursuit literature. In a demonstration with drones constrained to a fixed height and governed by the unicycle kinematics model, we demonstrate that direct sim-to-real transfer is possible.

A limitation of the current work is the need to train a network for each number of observable agents. This can be partially mitigated by using the same network and fixing the number of observable pursuers, as demonstrated in Section \ref{subsec:scalability}. Furthermore, it could be interesting to learn fixed-size state representation for neighboring agents by using deep sets~\cite{zaheer2017deep}, mean embeddings (similar to ~\cite{Huttenrauch2019}) or making use of Graph Neural Networks~\cite{zhou2019graph}. Other interesting directions of future work include exploring scenarios with numerous evaders, integrating smarter evader strategies by using the idea of safe-reachability ~\cite{7801073,ZHOU201664}, or implementing the evader as an RL agent and training both the evader and pursuer simultaneously similar to \cite{baker2020emergent}.

To enable more realistic applications, we also aim to extend the method from planar to 3D motion in the future. This will require careful consideration of the angular representation to avoid representational singularities. Furthermore, we aim to consider more realistic kinematic and perception models and include other constraints such as more unstructured environments with obstacles and varying arena sizes, a limited field of view, explore sim-to-real transfer further, and real-world applications to non-holonomic robots such as wheeled robots.

\bibliographystyle{IEEEtran}
\bibliography{refs}

\end{document}